\documentclass[11pt,a4paper,svgnames]{article}
\pdfoutput=1

\usepackage[utf8]{inputenc}
\usepackage{uniinput}
\usepackage{xcolor}
\usepackage{hyperref}
\hypersetup{colorlinks, breaklinks,
	urlcolor = DarkBlue,
	linkcolor = DarkBlue,
	citecolor = DarkBlue}
\usepackage{amsmath,amsfonts}
\usepackage{url}
\usepackage{bm}
\usepackage{mathtools}
\usepackage{relsize}
\usepackage{cancel}
\usepackage{csquotes}
\usepackage[plain]{fancyref}
\usepackage{subcaption}
\usepackage[margin=1in]{geometry}
\usepackage{mdframed}
\usepackage{cite}

\newcommand{\D}{\mathrm{d}}
\newcommand{\ie}{\textit{i.e.}}

\newcommand{\bea}{\begin{eqnarray}}
\newcommand{\eea}{\end{eqnarray}}
\newcommand{\be}{\begin{equation}}
\newcommand{\ee}{\end{equation}}

\usepackage[format=plain,
textfont=it
]{caption}
\usepackage{microtype}

\usepackage{graphicx}
\makeatletter
\newcommand*\biggercdot{{
		\hskip 8pt
		\color{Maroon}\mathpalette\bigcdot@{.5}
		\hskip 8pt
}}
\newcommand*\bigcdot{{
		\hskip 8pt
		\color{Maroon}\mathpalette\bigcdot@{.4}
		\hskip 8pt
}}
\newcommand*\bigcdot@[2]{\mathbin{\vcenter{\hbox{\scalebox{#2}{\color{Maroon}$\m@th#1\bullet$}}}}}
\makeatother

\newcommand{\Title}{Bubble needs strings}
\newcommand{\Authora}{Souvik Banerjee}
\newcommand{\Authorb}{Ulf Danielsson}
\newcommand{\Authorc}{Suvendu Giri}

\newcommand{\emaila}{souvik.banerjee\allowbreak@physik.uni-wuerzburg.de}
\newcommand{\emailb}{\{ulf.danielsson,suvendu.giri\}\allowbreak@physics.uu.se}

\begin{document}
    \vspace*{-40pt}
    {\hfill \texttt{UUITP-33/20}}
    \vskip 10pt
	\hskip -17pt{\huge \bfseries \Title}
	\vskip 15pt
	\hskip -15pt{\large \Authora $^\dagger$ $\bigcdot$\Authorb $^\star$ $\bigcdot$ \Authorc $^\star$}
	\vskip 10pt
	
	{\footnotesize $^\dagger$ Institut für Theoretische Physik und Astrophysik,
		Julius-Maximilians-Universität Würzburg,
		\vskip 1pt
		$~\,$ Am Hubland, 97074 Würzburg, Germany
		\vskip 2pt
		$^\star$ Institutionen för fysik och astronomi,
		Uppsala Universitet, Box 803, SE-751 08 Uppsala, Sweden
		\vskip 5pt
		\hskip 7pt\texttt{email: \emaila,\emailb }
	}
	\begin{center}
		\begin{minipage}[t]{\textwidth} 
		\begin{mdframed}[outerlinewidth=0pt,backgroundcolor=orange!10,linecolor=orange!10,innerleftmargin=10pt, innerrightmargin=10pt,innertopmargin=10pt,innerbottommargin=15pt]
		\small{
		\noindent \emph{In this paper, we want to emphasize the pivotal role played by strings in the model realizing de Sitter using an expanding bubble, proposed and subsequently developed in \cite{Banerjee:2018qey, Banerjee:2019fzz, Banerjee:2020wix}. Contrary to the Randall-Sundrum model of brane-localized gravity, we use the end points of radially stretched strings to obtain matter sourcing gravity induced on the bubble wall.  This allows us to reinterpret the possible volume divergence coming from naive dimensional reduction as mass renormalization in four dimensional particle physics. Furthermore, we argue that the residual time dependence in the bulk, pointed out by some recent work as a possible shortcoming of such models, is automatically cured in presence of these stringy sources.}	
		}
		\end{mdframed}
		\end{minipage}
	\end{center}
	\tableofcontents
    
    \section{Introduction}
    \label{sec:intro}
    
    To find de Sitter space in string theory has turned out to be a great challenge (see \cite{Danielsson:2018ztv,Cicoli:2018kdo} and references therein). The attempted constructions have been criticized, and conjectures put forward suggesting that string theory cannot accommodate de Sitter (see \cite{Brennan:2017rbf,Palti:2019pca} for a review). In view of this, it is important to find alternative approaches. Motivated by this, an attempt was made to construct a new stringy model of dark energy, with a four dimensional expanding universe on the surface of a spherical bubble expanding in five dimensional AdS space in \cite{Banerjee:2018qey}; this was called the \emph{shellworld}. End points of strings stretching out in the fifth direction -- in which the shell expands -- give rise to four dimensional matter on the shellworld. The role of the stretched strings, and the nature of four dimensional gravity, was subsequently elaborated in \cite{Banerjee:2019fzz,Banerjee:2020wix}. By computing the bounce action of the shell in the presence of a cloud of strings and a Schwarzschild mass in five dimensions, \cite{Koga:2019yzj} showed that the preferred nucleation channel is a bubble with slightly sub-critical tension, giving a phenomenologically realistic four dimensional cosmological constant. 
    Recently, a top down construction of a $(p+1)$ dimensional dS shellworld, via the nucleation of a $p$ dimensional bubble in an AdS$_{p+2}\times \mathbb{S}^q$ bulk, was done for $p=1,5$ in ten dimensional non-supersymmetric string theory by \cite{Basile:2020mpt}. 
     
    In this paper, we want to highlight and further explain how four dimensional gravity arises in this construction -- in particular the value of the four dimensional Planck scale.
    In order to obtain an effective four dimensional theory from a higher dimensional theory, such as string theory, the extra dimensions need to be inaccessible to processes occurring at energy scales of the four dimensional theory. There are a few ways in which this can happen. 
    
    The simplest and most common approach is to use dimensional reduction à la Kaluza and Klein. When the extra dimensions are compact and small, quantization forces any non-zero momentum along the extra dimensions to be large. These are interpreted as heavy Kaluza-Klein modes from the point of view of the four dimensional theory. 
    A particle with no momentum in the extra dimensions is delocalized over those small compact dimensions, and its motion is fully described by equations of motion coming from the effective four dimensional theory. Furthermore, for a compactification from $d$ dimensions to four, with a
    compact space of volume $\mathcal{V}_{d-4}$, 
    the four dimensional Newton's constant is given by 
    \begin{equation}
    G_4 = \frac{G_d}{\mathcal{V}_{d-4}}.
    \end{equation}
    Therefore, the strength of gravitational interactions in four dimensions, $G_4$, falls off as the size of the extra dimensions grows.
    
    Another possibility was proposed by Randall and Sundrum \cite{Randall:1999vf}, by making use of a \emph{braneworld}. In this case, the total space is a warped product of the braneworld and the extra dimensions. Due to this non-trivial warping, the four dimensional Planck scale turns out to be finite, despite the extra dimensions being as large as desired. An important difference compared to the \emph{unwarped} Kaluza-Klein reduction above, is that the warped extra dimensions restrict the wave functions from spreading out into the extra dimensions. Concretely, their construction consists of a five dimensional AdS space viewed as a warped product of the fifth dimension over a four dimensional Minkowski space (corresponding to the braneworld). Matter, and even gravity are localized on the braneworld. For an AdS$_5$ of curvature $k$, 
    the four dimensional Newton's constant is given by
    \begin{equation}
    G_4 = G_5 k.
    \end{equation}
    In contrast to the Kaluza-Klein reduction above, the four dimensional Newton's constant depends only on the curvature, rather than the size of the extra dimension.
    
    Our shellworld proposal in \cite{Banerjee:2018qey,Banerjee:2019fzz,Banerjee:2020wix} differs from both of the above. Like in the Randall-Sundrum (RS) scenario, we make use of a braneworld. However, our shellworld is a co-dimension one bubble that nucleates in a metastable AdS$_5$ space and expands outwards, causing a phase transition to a more stable AdS$_5$ vacuum. This makes the geometry across it \emph{inside-outside} rather than the \emph{inside-inside} of the RS construction.\footnote{Inside (outside) refers to direction in which radial slices decrease (increase) in volume as one moves away from the brane.} Once a bubble with the right tension is nucleated, its expansion will be such that a dS geometry with an effective dark energy is induced on the shellworld. 
    As shown in \cite{Banerjee:2018qey}, in presence of strings stretching out in the extra dimensions, one recovers the full four dimensional Friedmann's equations on the brane with an effective four dimensional Newton's constant
    \begin{equation}
    \label{eq:G4-G5-intro}
    G_4 = 2G_5 \left(\frac{1}{k_+}-\frac{1}{k_-}\right)^{-1},
    \end{equation}
    where $k_\pm$ are the curvatures of the AdS$_5$ spaces outside and inside of the bubble respectively. Such a nucleation event requires $k_- > k_+$.
    Additionally, matter particles in the shellworld can be realized as the end points of strings on the brane.
    
    In this work, we will present a comparative study of these three approaches. While in the case of inside-inside one gets a finite effective theory of gravity, a naive dimensional reduction in the case of inside-outside yields a divergent result. This divergence is a very natural consequence of having a divergent spacetime volume  outside the bubble. Naively, this pushes $G_4$ to zero.
    This is in sharp contrast to the result of \fref{eq:G4-G5-intro}, which we obtained for our model, and the reason for this is the following. Unlike the RS scenario, our sources are not localized on the brane but extend as strings in the extra dimension. This is also different from a naive Kaluza-Klein reduction, where the source is a point particle in four dimensions with its wave function smeared over the extra dimensions. 
    
    To make this difference explicit, we will study the motion of a probe string in the background of a heavy stationary string, with its endpoint on the shellworld. We will find that while the probe string remains radially stretched as it is gravitationally attracted towards the source string, its end point follows a four dimensional geodesic on the shellworld. As a consequence, even though the proper distance between the strings increases as we move away from their end points, the effective gravitational force between them increases to compensate for this. Connecting to the dimensional reduction argument, and the associated volume divergence, this very simple bulk dynamics can be interpreted as a mass renormalization scheme on the moving brane. We will also point out how this very way of interpreting this dynamics is consistent with the holographic renormalization procedure in light of the AdS/CFT correspondence.
    
    We will further see that this is precisely the dynamics in the bulk that saves us from the undesirable time dependence, which was a concern in some recent papers \cite{Montefalcone:2020vlu, Karch:2020iit}. In the context of a $2$-brane model, \cite{Karch:2020iit} showed that the effective $G_4$ depends on the time dependent relative position (parametrized by the {\it radion}) of a cutoff brane. They did find that a change of frame to Einstein frame actually makes $G_4$ time independent, but this was of no help since any matter dependent on the radion would have the undesired time dependence. A very similar time dependence of the matter part also shows up in \cite{Montefalcone:2020vlu}. 
    
    In our case, there is no such cut-off brane in the inside region. We do not need a second brane simply because matter is introduced in a completely different way, using the strings that stretch outwards as far as one likes. Even if they eventually end on another brane on the outside, this will still not introduce any further time dependence. The motion of the strings keeps $G_4$ independent of the cut-off, with the cut-off dependence taken care of by a mass renormalization of the particles induced on the braneworld. The particles themselves, simply follow the geodesics in the curved space time, ignorant of what is happening in the bulk, in particular, how long the strings are. This  picture will not break down until expansion makes the branes come sufficiently close.
    
    The rest of the paper is organized as follows. In \fref{sec:in-out}, we generalize the RS construction to include the inside-outside shellworld construction, highlighting some of the major differences that leads to a different four dimensional Newton's constant. In \fref{sec:strings}, we examine the role of stretched strings in the shellworld construction. We perform a worldsheet analysis to understand their role in producing four dimensional gravity on the shellworld, and comment on the analogy between mass renormalization and holographic renormalization. In \fref{sec:time-dep}, we comment on the time dependence that was pointed out in recent results of \cite{Montefalcone:2020vlu,Karch:2020iit} and how it is evaded in the shellworld construction. Finally, we conclude with a summary in \fref{sec:discussion}.
    
    \section{Inside, outside}
    \label{sec:in-out}
    We will start with a general warped AdS geometry arising from the five dimensional action
    \be
    S = \frac{1}{16 \pi G_5}\int \D ^5 x \sqrt{-g^{\text (5)}} \left(R - 2 \Lambda + \mathcal{L}_m\right) .
    \label{EH+m}
    \ee 
    \noindent Here $G_5$ is the five dimensional Newton's constant and $g^{\text (5)}_{MN}$ is the five dimensional bulk metric. $\mathcal{L}_m$ denotes the matter Lagrangian giving rise to a five dimensional stress-energy tensor
    \be
    T_{MN} \coloneqq -\frac{2}{\sqrt{-g^{\text (5)}} } \frac{\delta \left(\sqrt{-g^{(5)}}\mathcal{L}_m\right)}{\delta g^{{\text (5)}MN}}, 
    \ee
    where ${M,N} = 1, \ldots,5$. While evaluating this stress-energy tensor, we fix $\delta g_{zz} = 0$ throughout. This ensures that we are only considering fluctuations on the shell. 
    The five dimensional Einstein's equations read 
    \be
    R_{MN} - \frac{1}{2} R g_{MN} + \Lambda g_{MN} = 8 \pi G_5 T_{MN} . 
    \label{eq:EE1}
    \ee 
    %
	Taking an ansatz for the metric of the form
    \be
    \D s^2 = e^{2 A(z)} \eta_{\mu\nu} \D x^\mu \D x^\nu + \D z^2, 
    \label{eq:AdS-ansatz}
    \ee
    where ${\mu,\nu} = 1, \ldots,4$,
    yields the following equations for the function $A(z)$:
    \begin{equation}
    A'(z)^2 = \frac{1}{6} \left(8 \pi G_5 T^{z}_z - \Lambda\right), \qquad
    A''(z) = \frac{8\pi G_5}{3}  \left(T^{t}_t - T^z_z \right).
    \label{eq:eqnA}
    \end{equation}
    %
    When $\mathcal{L}_m = 0$, the solution to the above equation is an empty AdS$_5$ spacetime with a negative cosmological constant $\Lambda < 0$:
    \be
    \label{eq:scale-factor}
    A = \pm \sqrt{\frac{-\Lambda}{6}} z + c_0,
    \ee
    $c_0$ being an integration constant that can be absorbed in the metric of \fref{eq:AdS-ansatz} through a redefinition of the transverse coordinates, $x_\mu$.
    Defining the AdS$_5$ curvature $k = \sqrt{-\Lambda/6}$, the metric of \fref{eq:AdS-ansatz} becomes
    \begin{equation}
    \D s^2 = e^{\pm 2kz}\eta_{\mu\nu} \D x^\mu \D x^\nu + \D z^2,
    \end{equation}
    where the $z$ coordinate increases from $z\rightarrow-\infty$ at the center of AdS to $z \rightarrow +\infty$ at the boundary.

    
    \subsection{A generalized braneworld}
    
    Let us consider a generalization of the RS construction, where a brane is placed at $z=0$ and separates two distinct AdS$_5$ spaces across it. Unlike RS, we do not require a $\mathbb{Z}_2$ symmetry across the brane. Denoting the corresponding AdS curvatures on either side of the brane by $k_\pm$, the metric of the entire space is given by \fref{eq:AdS-ansatz} with 
    \begin{equation}
    A(z) = \epsilon_+ k_+ \Theta(z) z + \epsilon_- k_- \Theta(-z) z,
    \end{equation}
    where $\epsilon_\pm = \pm 1$ and $z$ is the universal fifth direction that spans from $z \rightarrow -\infty$ at the center of AdS$^-_5$ to $z \rightarrow + \infty$ at the boundary of AdS$^+_5$. The five dimensional metric is continuous across the brane, but the derivatives suffer a discontinuity. Evaluated on the brane, this gives $A''(0)=\left(\epsilon_+k_+ - \epsilon_- k_-\right)\delta(0)$, which together with $T^z_z=0$ requires a localized stress tensor on the brane given by \fref{eq:eqnA}
    \begin{equation}
    T^M_N = \frac{3}{8 \pi G_5} \left(\epsilon_+ k_+  - \epsilon_- k_-\right) \delta(z)\, \delta^\mu_\nu \, \delta^M_\mu \, \delta^\nu_N.
    \end{equation}
    The generalized braneworld scenario therefore requires the above stress tensor on the brane together with the five dimensional metric:
    \begin{equation}\label{eq:pmmetric}
    \D s^2 = e^{2z\left(\epsilon_+ k_+ \Theta(z) + \epsilon_- k_- \Theta(-z)\right)} \eta_{\mu \nu}\D x^\mu \D x^\nu + \D z^2.
    \end{equation}
    Let us now specialize to the case of interest -- shellworlds. In this case, the bulk spacetime before nucleation of the brane is AdS$^+_5$, implying that $A(z)_\textrm{before}=k_+z$, giving an exponentially increasing warp factor. After the nucleation event, the spacetime outside the bubble remains unchanged, while the interior of the bubble is replaced by AdS$^-_5$, with a lower and more negative cosmological constant. Since the interior looks like an empty AdS space, the warp factor increases from zero at the center of the bubble, to a finite value on the bubble. This corresponds to $\epsilon_-=\epsilon_+=+1$ in \fref{eq:pmmetric}. The RS-scenario is different. There the brane cuts off the AdS$_5$ spacetime and has a $\mathbb{Z}_2$ symmetry across it, giving a finite volume on both sides. The warp factor therefore increases from the center of AdS to the brane and then symmetrically goes to zero away from the brane. 
    \begin{figure}[t]
        \begin{subfigure}[b]{0.49\textwidth}
            \def\svgwidth{\linewidth}
\begingroup%
  \makeatletter%
  \providecommand\color[2][]{%
    \errmessage{(Inkscape) Color is used for the text in Inkscape, but the package 'color.sty' is not loaded}%
    \renewcommand\color[2][]{}%
  }%
  \providecommand\transparent[1]{%
    \errmessage{(Inkscape) Transparency is used (non-zero) for the text in Inkscape, but the package 'transparent.sty' is not loaded}%
    \renewcommand\transparent[1]{}%
  }%
  \providecommand\rotatebox[2]{#2}%
  \newcommand*\fsize{\dimexpr\f@size pt\relax}%
  \newcommand*\lineheight[1]{\fontsize{\fsize}{#1\fsize}\selectfont}%
  \ifx\svgwidth\undefined%
    \setlength{\unitlength}{349.00719096bp}%
    \ifx\svgscale\undefined%
      \relax%
    \else%
      \setlength{\unitlength}{\unitlength * \real{\svgscale}}%
    \fi%
  \else%
    \setlength{\unitlength}{\svgwidth}%
  \fi%
  \global\let\svgwidth\undefined%
  \global\let\svgscale\undefined%
  \makeatother%
  \begin{picture}(1,0.65265455)%
    \lineheight{1}%
    \setlength\tabcolsep{0pt}%
    \put(0.22841058,0.33239031){\color[rgb]{0,0,0}\makebox(0,0)[t]{\lineheight{1.25}\smash{\begin{tabular}[t]{c}in\end{tabular}}}}%
    \put(0.73465407,0.33239031){\color[rgb]{0,0,0}\makebox(0,0)[t]{\lineheight{1.25}\smash{\begin{tabular}[t]{c}out\end{tabular}}}}%
    \put(0,0){\includegraphics[width=\unitlength,page=1]{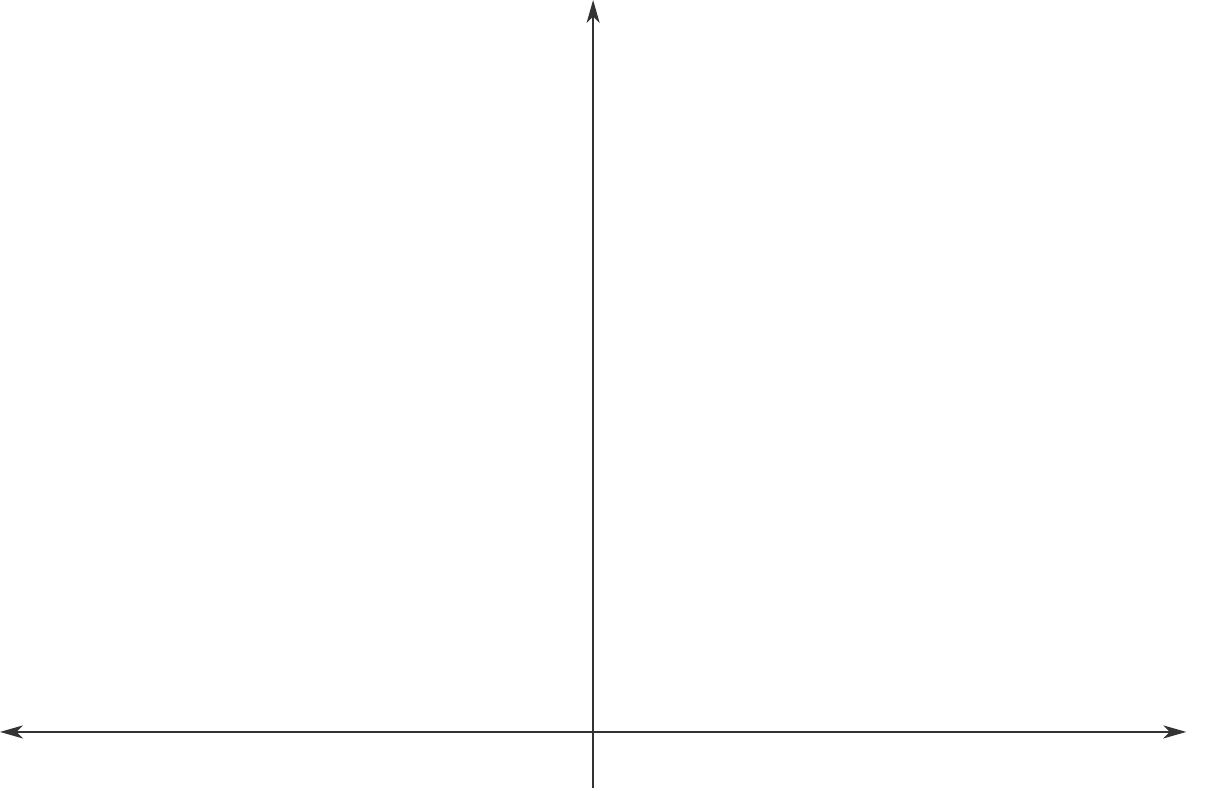}}%
    \put(0.97195449,0.0050534){\color[rgb]{0,0,0}\makebox(0,0)[t]{\lineheight{1.25}\smash{\begin{tabular}[t]{c}$z$\end{tabular}}}}%
    \put(0.55931269,0.59946349){\color[rgb]{0,0,0}\makebox(0,0)[t]{\lineheight{1.25}\smash{\begin{tabular}[t]{c}$e^{A(z)}$\end{tabular}}}}%
    \put(0,0){\includegraphics[width=\unitlength,page=2]{shell_warp.pdf}}%
  \end{picture}%
\endgroup%

            \caption{Warp factor for the shellworld: $\epsilon_-=\epsilon_+=+1$}
            \label{fig:shell_warp}
        \end{subfigure}
        \begin{subfigure}[b]{0.49\textwidth}
            \def\svgwidth{\linewidth}
\begingroup%
  \makeatletter%
  \providecommand\color[2][]{%
    \errmessage{(Inkscape) Color is used for the text in Inkscape, but the package 'color.sty' is not loaded}%
    \renewcommand\color[2][]{}%
  }%
  \providecommand\transparent[1]{%
    \errmessage{(Inkscape) Transparency is used (non-zero) for the text in Inkscape, but the package 'transparent.sty' is not loaded}%
    \renewcommand\transparent[1]{}%
  }%
  \providecommand\rotatebox[2]{#2}%
  \newcommand*\fsize{\dimexpr\f@size pt\relax}%
  \newcommand*\lineheight[1]{\fontsize{\fsize}{#1\fsize}\selectfont}%
  \ifx\svgwidth\undefined%
    \setlength{\unitlength}{349.00719096bp}%
    \ifx\svgscale\undefined%
      \relax%
    \else%
      \setlength{\unitlength}{\unitlength * \real{\svgscale}}%
    \fi%
  \else%
    \setlength{\unitlength}{\svgwidth}%
  \fi%
  \global\let\svgwidth\undefined%
  \global\let\svgscale\undefined%
  \makeatother%
  \begin{picture}(1,0.65265455)%
    \lineheight{1}%
    \setlength\tabcolsep{0pt}%
    \put(0.23577583,0.33668822){\color[rgb]{0,0,0}\makebox(0,0)[t]{\lineheight{1.25}\smash{\begin{tabular}[t]{c}in\end{tabular}}}}%
    \put(0.74201933,0.33668822){\color[rgb]{0,0,0}\makebox(0,0)[t]{\lineheight{1.25}\smash{\begin{tabular}[t]{c}in\end{tabular}}}}%
    \put(0,0){\includegraphics[width=\unitlength,page=1]{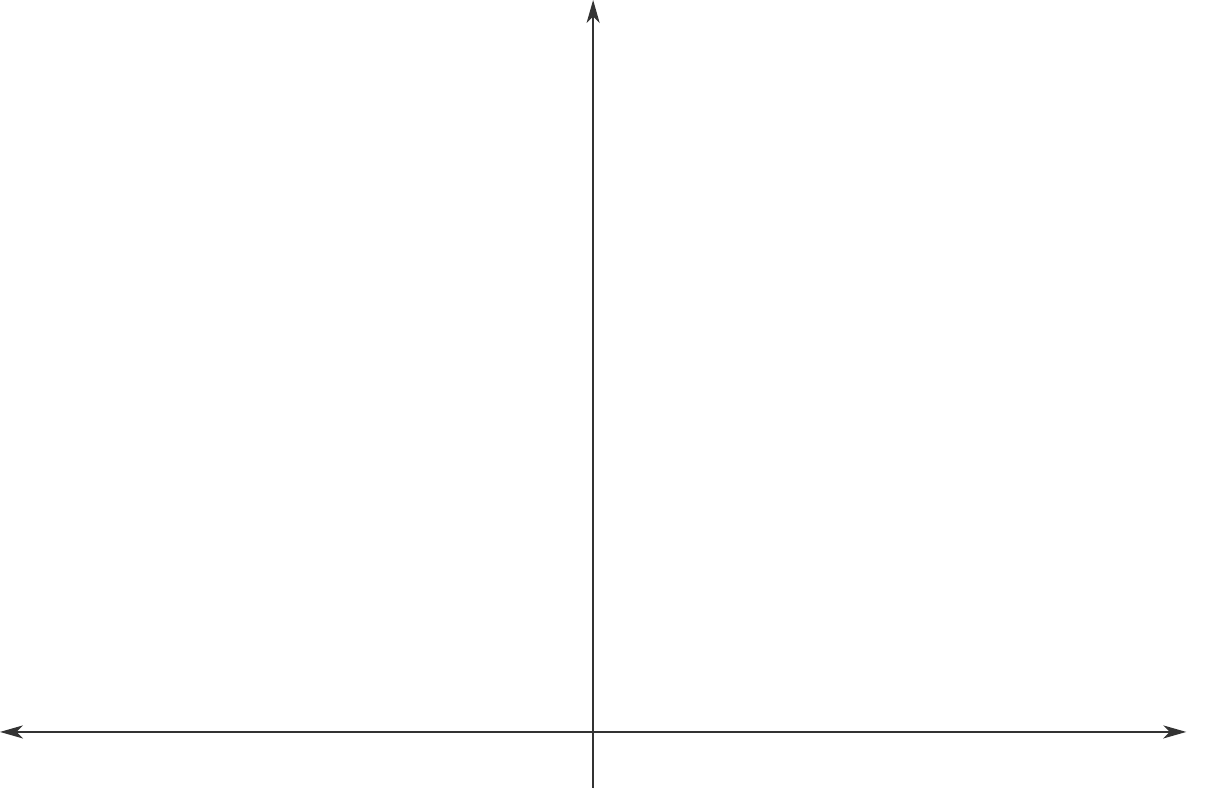}}%
    \put(0.97195449,0.0050534){\color[rgb]{0,0,0}\makebox(0,0)[t]{\lineheight{1.25}\smash{\begin{tabular}[t]{c}$z$\end{tabular}}}}%
    \put(0.55931269,0.59946349){\color[rgb]{0,0,0}\makebox(0,0)[t]{\lineheight{1.25}\smash{\begin{tabular}[t]{c}$e^{A(z)}$\end{tabular}}}}%
    \put(0,0){\includegraphics[width=\unitlength,page=2]{rs_warp.pdf}}%
  \end{picture}%
\endgroup%

            \caption{Warp factor for RS: $\epsilon_-=+1,\epsilon_+=-1$}
            \label{fig:rs_warp}
        \end{subfigure}
        \caption{Warp factors for the shellworld vs the RS setup. The shellworld construction has an inside and an outside, as can be seen from the increasing warp factor beyond the position of the shell at $z=0$. The RS construction, on the other hand, has insides on both sides of the brane. Additionally, while the AdS spaces necessarily have different curvatures across the brane in the shellworld construction, the RS scenario has a $\mathbb{Z}_2$ symmetry across the brane. The above plots are for $k_-=3,k_+=1$, and $k_\textrm{\scshape rs}=3$.}
        \label{fig:warp_factors}
    \end{figure}
    This corresponds to the choice of $\epsilon_-=+1$ and $\epsilon_+=-1$. This gives the familiar RS metric:
    \be
    \D s_{\textrm{\scshape rs}}^2 = e^{-2 k |z|} \eta_{\mu\nu} \D x^\mu \D x^\nu + \D z^2.
    \label{RS-metric}
    \ee
    The warp factors for the RS and the shellworld are plotted in \fref{fig:warp_factors}.

    
    
    
    
    

    \subsection{Effective gravity from dimensional reduction}
    
    Let us dimensionally reduce the five dimensional Einstein-Hilbert action in the absence of matter sources extending into the extra dimension. Starting from the action
    \be
    S_5 \supset \frac{1}{16 \pi G_5}\int \D ^5 x \sqrt{-g^{\text {(5)}}} R^{\text{(5)}},
    \label{eq:R-EH}
    \ee
    and plugging in the metric of \fref{eq:pmmetric}, we obtain, through simple dimensional reduction
    \begin{eqnarray}
    \sqrt{-g^{\text {(5)}}} &=&  e^{4 \left(\epsilon_+ k_+ \Theta(z) +   \epsilon_- k_- \Theta(-z)\right) z} \sqrt{-g^{\text {(4)}}} + \cdots \nonumber \\
    R^{\text{(5)}} &=&  e^{-2 \left(\epsilon_+ k_+ \Theta(z) +   \epsilon_- k_- \Theta(-z)\right) z} R^{\text{(4)}} + \cdots
    \label{dim-red}
    \end{eqnarray}
    This gives
    \begin{eqnarray}
    S_5&\supset& \frac{1}{16 \pi G_5}\int \D ^4 x \sqrt{-g^{\text {(4)}}} R^{\text{(4)}} \int_{-\infty}^{\infty} \D z \, e^{2 \left(\epsilon_+ k_+ \Theta(z) +  \epsilon_- k_- \Theta(-z)\right) z} \nonumber \\
    &\equiv& \frac{1}{16 \pi G_4}\int \D^4 x \sqrt{-g^{\text {(4)}}} R^{\text{(4)}},
    \label{R-EH1}
    \end{eqnarray}
    where the effective $4$ dimensional Newton's constant, $G_4$ in the last line of \eqref{R-EH1} is defined as
    \be
    G_5 = G_4 \int_{-\infty}^{\infty} \D z \, e^{2 \left(\epsilon_+ k_+ \Theta(z) +   \epsilon_- k_- \Theta(-z)\right) z}  .
    \label{eq:G45-G}
    \ee
    \\
    For the RS scenario ($\epsilon_+ = - \epsilon_- = -1, k_+ = k_- = k$) we get
    \be
    G_5 = \frac{1}{k} G_4   .
    \label{eq:G45-RS}
    \ee
    The scenario we are going to consider in the rest of the paper has one ``exterior'' AdS with cosmological constant $\Lambda_+$ (so that $\epsilon_+ = 1$) and one ``interior'' AdS with cosmological constant $\Lambda_-$ (so that $\epsilon_- = -1$) separated by the brane.   In this case, the dimensional reduction fails to produce a finite $G_4$.  This is a consequence of the fact that we have a diverging scale factor corresponding to an infinitely extended extra dimension. This is unlike the RS case, where the finiteness is guaranteed through the warping of the extra dimension. 
    
    In order to make the origin of the divergence explicit, let us separate the divergent piece by evaluating the integral on the right hand side of \fref{eq:G45-G} with a hard cut off, $z_0$, in the radial direction
    \be
    G_5 = G_4 \int_{-\infty}^{z_0} \D z \, e^{2 \left(\epsilon_+ k_+ \Theta(z) +   \epsilon_- k_- \Theta(-z)\right) z}.
    \label{eq:G45-G1}
    \ee
    This yields
    \be
    G_5 = \frac{G_4 }{2} \left(\frac{1}{\epsilon_- k_-} - \frac{1}{\epsilon_+ k_+}\right) + G_4 \, \frac{1}{2 \epsilon_+ k_+} e^{2 \epsilon_+ k_+ z_0}.
    \label{eq:G45-G2}
    \ee
    There are seemingly two problems with this expression. The first and most obvious problem, is that the second term diverges for inside-outside. This is different from inside-inside, where this term is automatically zero by virtue of the warping. Since the second term merely reflects the volume divergences arising out of infinite extra dimensions, it is natural to expect that this could be taken care of by a suitable renormalization scheme. We will see how this works below. The second problem is that the finite term, namely the first term in \fref{eq:G45-G2}, is negative for $k_ - > k_+$. As we will explain, this is actually the correct sign.

    \section{A nucleating bubble and few hanging strings}
    \label{sec:strings}
    
    Let us come back to what we started with - our shellworld dynamically realized as a true AdS$_5$ bubble nucleating and eventually expanding inside a false AdS$_5$ vacuum. This was the model proposed in \cite{Banerjee:2018qey} and then developed in \cite{Banerjee:2019fzz} and in \cite{Banerjee:2020wix}.
    
    \subsection{Effective gravity -- \`{a} la Friedmann}
    
    Consider an effective five dimensional theory with two non-degenerate AdS$_5$ vacua having cosmological constants $\Lambda_\pm = -6/L_\pm^2 = -6 k_\pm^2$, with $k_- > k_+$. Assume, in addition, that the five dimensional cosmological constant is supported by fluxes. In this theory, there exist spherical instantons corresponding to branes charged under the form field sourcing the flux.\footnote{It would be interesting to see if other instabilities, apart from the ones arising due to flux tunneling as discussed here, could also give rise to analogous shellworlds. Bubble of nothing instabilities of non-supersymmetric vacua, for example, as recently discussed in \cite{Dibitetto:2020csn, GarciaEtxebarria:2020xsr} could be one such possibility.} 
    The metastable AdS$_5^+$ vacuum in this theory would undergo a first-order phase transition to AdS$_5^-$ via the nucleation of a spherical brane containing true vacuum inside of it. This bubble nucleates at rest, and accelerates outward towards the boundary of AdS$_5^+$. Choosing coordinates such that the center of AdS$_5^+$ is the center of the nucleated bubble (and hence also the center of AdS$_5^-$), the spacetime after nucleation can be written as
    \begin{equation}
    \D s^2 = -f(r)\D t^2 + f(r)^{-1}\D r^2 + r^2 \D \Omega_3^2,
    \end{equation}
    with
    \begin{equation}
    f(r)=\left(1+k_-^2r^2\right) + \Theta\left(r-a(t)\right)\left(k_+^2r^2-k_-^2r^2\right),
    \end{equation}
    where $r=a(t)$ is the instantaneous radius of the spherical shell, and $\Theta$ is the Heaviside theta function.
    In terms of proper time on the shell $\tau$, the induced metric takes the Friedmann-Lema\^itre-Robertson-Walker (FLRW) form $\D s^2 = -\D \tau^2 + a(\tau)^2\D \Omega_3^2$. Einstein's equations on the shell are given by the thin-shell junction conditions \cite{Israel:1966rt,doi:10.1002/andp.19243791403,doi:10.1002/andp.19243780505}, which determine its expansion in terms of the tension of the shell. The $\tau\tau$ component gives,
    \begin{equation}
    \sigma = \frac{3}{8\pi G_5} \left(\sqrt{k_-^2 + \frac{1+\dot{a}^2}{a^2}}-\sqrt{k_+^2 + \frac{1+\dot{a}^2}{a^2}}\right).
    \end{equation}
    In the limit of large curvatures $k_\pm$, this can be rewritten as the four dimensional Friedmann equations on the shell
    \begin{equation}
    H^2 \equiv \frac{\dot{a}^2}{a^2} = -\frac{1}{a^2} + 2k_- k_+ - \frac{8\pi G_5 \sigma}{3}\left(\frac{2k_-k_+}{k_- - k_+}\right).
    \end{equation}
    The first term on the right indicates a spatially closed universe, while the last two terms are proportional to the effective four dimensional cosmological constant. A Minkowski universe corresponds to this term being zero, which requires the shell to have a critical tension $\sigma_\textrm{crit} \coloneqq 3\left(k_- - k_+\right)/(8\pi G_5)$. For a shell with slightly sub-critical tension\footnote{In the RS-scenario one needs a {\it supercritical} tension to get a dS- universe. See figure 4.} $\sigma = \sigma_\textrm{crit}\left(1-\epsilon\right)$, with $1 \gg \epsilon >0$, this gives
    \begin{equation}
    \frac{\dot{a}^2}{a^2} = -\frac{1}{a^2} + 2k_- k_+ \epsilon + \mathcal{O}\left(\epsilon^2\right),
    \end{equation}
    which corresponds to an empty dS universe. Generalizing the five dimensional spacetime to AdS-Schwarzschild induces radiation on the shellworld, while adding radially stretched strings (resulting in a cloud of strings) gives rise to four dimensional matter.\footnote{The strings that we consider here are unbreakable strings. While such strings are themselves charged under gauge fields, we consider neutral bound states which can give rise to neutral matter in four dimensions.} These contributions can be read off from the resulting Friedmann equation, and the strength of these interactions gives the four dimensional Planck scale:
    \begin{equation}\label{eq:friedmann}
    \frac{\dot{a}^2}{a^2} = -\frac{1}{a^2} + \frac{8\pi G_4}{3} \left[\Lambda_4 + \frac{1}{2\pi^2a^4} \left(\frac{M_+}{k_+}-\frac{M_-}{k_-}\right) + \frac{3}{8\pi a^3} \left(\frac{\alpha_+}{k_+}-\frac{\alpha_-}{k_-}\right)  \right]
    + \mathcal{O}\left(\epsilon^2\right),
    \end{equation}
    where $M_\pm, \alpha_\pm$ are the Schwarzschild masses and the tension of the cloud of strings respectively, and the four dimensional scales can be read off as:
    \begin{equation}
    \label{kappa4}
    G_4 \equiv \frac{2k_-k_+}{k_- - k_+}G_5, \qquad \textrm{and} \qquad
    \Lambda_4 \equiv \sigma_\textrm{crit}-\sigma = \epsilon \sigma_\textrm{crit}.
    \end{equation}
    The mass parameter $M_+$ is related to $M_-$ by the total mass-energy of the shell (its tension $σ$ as well as the kinetic energy of expansion) as it must, from energy conservation in a Brown-Teitelboim nucleation.
    
    \subsection{Divergence revisited: mass renormalization}
    \label{sec:mass-renorm}
    As mentioned  already in the introduction, with the non-localized string sources, our model is very different from what we used for the naive dimensional reduction, where the source is assumed to be an effective point particle with normalizable wavefunction. As we will explain below, the divergence that we found turns out to be just an artifact of this difference.
    
    To understand what happens, let us assume a stationary heavy source in the form of a string with its endpoint on the brane world. Then consider another string acting as a probe -- both stretching out radially.
    As the probe moves in the metric caused by the non-local stringy source, it remains radially stretched. The end point on the brane follows a four dimensional geodesic in the induced metric, while the rest of the string moves in sync. This means, even though the proper distance between the strings increases as we move away from the brane world, the effective gravitational force increases as one would expect from an effective four dimensional viewpoint. 
    
    The simplest way to see this, is to use the worldsheet action of the string.
    The background created by a radially stretched string was computed perturbatively in \cite{Banerjee:2020wix}. It is given by
    \begin{equation}
    \D s^2 = \D z^2 + a(z)^2 \left[-f_1(r) \D t^2 + f_2(r) \D r^2 + f_3(r) r^2 \D \Omega_2^2\right],
    \end{equation}
    where at linear order in the tension of the string $\chi$, \footnote{It was shown in \cite{Banerjee:2019fzz} that the \enquote{average tension density} $\alpha$ in \fref{eq:friedmann} is related to the tension of an individual string $χ$ via the number density $N/V_3$ as $\alpha \sim χ N/V_3$. }
    \begin{equation}\label{eq:f123linearorder}
    f_1(r)=\left(1-\frac{4\chi}{r}\right),\quad f_2(r)=\left(1+\frac{2\chi}{r}\right),\quad f_3(r)=\left(1+\frac{\chi}{r}\right),\quad a(z)=\exp \left(kz\right).
    \end{equation}
    This is asymptotically AdS$_5$ as expected.
    The worldsheet action of a probe string in this background is given by the Nambu-Goto action. To compute it, let us choose the static gauge, $t=\tau$, and use the reparametrization invariance of the worldsheet to choose $\sigma=z$. A string stretching in the $r,z$ direction can be embedded as $X^\mu=\left(t,r(t,z),z\right) $. The string stretches between the shellworld on one side ($z=0$) and a cutoff brane on the other side ($z=z_\textrm{cutoff}$), giving Dirichlet boundary conditions on both ends $\partial r(t,z_\textrm{cutoff})/\partial z=\partial r(t,0)/\partial z=0$. With this embedding, the Nambu-Goto action (for a probe string of tension $T$) reads
    \begin{equation}\label{eq:NG_action}
    \begin{split}
    S_\textrm{\scshape ng} 
    &= -T \int \D t~\D z~ a(z) \sqrt{\left(\dot{X}\cdot X^\prime\right)^2 - \dot{X}^2 \left(1+a(z)^2X^\prime{}^2\right)}\\
    &= -T \int \D t~\D z~
    e^{kz}\sqrt{f_1(r)-f_2(r)\dot{r}^2 + e^{2kz} f_1(r)f_2(r)r^\prime{}^2},
    \end{split}
    \end{equation}
    where primes and dots represent derivatives with respect to $z$ and $t$ respectively, and the metric coefficients are given by \fref{eq:f123linearorder}. For a radially stretched string ($r^\prime=0$) the linearized Euler-Lagrange equation becomes
    \begin{equation}\label{eq:geodesic_equation_1}
    \ddot{r} = \frac{\chi}{r^2}\left(5 \dot{r}^2 -2\right).
    \end{equation}
    This is simply the geodesic equation in a four dimensional Schwarzschild spacetime. As discussed in \cite{Banerjee:2020wix}, the endpoint of the string pulls on the brane and it is no longer situated at a fixed $z$. There is, however, a gauge freedom \cite{Garriga:1999yh,Giddings:2000mu} which can be used to change coordinates to make the brane appear flat (\ie, located at a fixed $z$). It was shown that in these coordinates, $f_1(r)=\left(1-3\chi/r\right), f_2(r)=\left(1+3\chi/r\right)$, and the induced metric on the brane at $z=0$ is Schwarzschild. With this metric, the equation of motion following from \fref{eq:NG_action} becomes
    \begin{equation}
    \ddot{r}=\frac{3\chi}{2r^2}\left(\dot{r}^2-1\right),
    \end{equation}
    which is the geodesic equation for a particle moving in a four dimensional Schwarzschild spacetime of mass $M=3\chi/2$, consistent with the claim in \cite{Banerjee:2020wix}.
    Therefore, a radially stretched string remains radial as it is attracted towards the heavy string. 
    The reason why it  stays radial and does not tilt in the $r-z$ plane can be seen from the Nambu-Goto action of \fref{eq:NG_action}. The potential energy associated to a particular configuration of the string can be easily read off from \fref{eq:NG_action} for a static configuration. For a radial string, this is given by
    \begin{equation}
    V_\textrm{radial} = T \int_0^{z_\textrm{cutoff}} \D z~a(z)\sqrt{f_1(r)},
    \end{equation}
    while for a non-radial string, it has an additional contribution which scales with $a(z)^2$,
    \begin{equation}
    V_\textrm{non-radial} = T \int_0^{z_\textrm{cutoff}} \D z~a(z)\sqrt{f_1(r)+ a(z)^2 f_1(r)f_2(r)r^\prime{}^2}.
    \end{equation}
    Any non-radial tilt, therefore, comes with an enormous 
    energy cost and such modes will not be excited in low energy processes. To linear order in $\chi$, we have
    \begin{equation}
    V_\textrm{non-radial} = T \int_0^{z_\textrm{cutoff}} \D z ~\frac{a(z)}{\sqrt{1+a(z)^2r^\prime{}^2}}\left[ \left(1-\frac{2\chi}{r}\right) + a(z)^2r^\prime{}^2 \left(1-\frac{\chi}{r}\right) \right] + \mathcal{O}\left(\chi^2\right).
    \end{equation}
    It is clear form above, that the stringy sources giving rise to non-normalizable modes in the bulk, play a pivotal role in giving the correct behavior of the gravitational potential. With this, one can also give an interpretation of the regularization scheme of $G_4$ in terms of a renormalization of the bare masses of the effective particles on the braneworld. As one moves away from the brane along the strings, the proper distance between the strings increases. However, as argued in the previous discussion, the strings remain straight with their endpoints in the form of four dimensional particles exhibiting the usual geodesic motion on the curved brane. The fact that the strength of the gravitational force remains constant between segments of the strings as we move away from the brane, even though the proper distance is increasing, is explained by the non-normalizable modes. If we instead consider the motion of the brane along the strings as the universe expands, we conclude that the strength of the gravitational force remains {\it the same} as a function of the proper distance in terms of the four dimensional theory on the brane. This is in the same spirit as in {\cite{Banerjee:2019fzz}, where we argued that the brane \emph{eats up} the strings and in this way provides the necessary energy for a particle to be carried upwards in the gravitational potential of the AdS-throat by the expansion. In case of strings that are so far away from each other that their mutual gravitational attraction can be ignored, the proper distance between the end points will increase as the brane eats strings and the universe expands. This is illustrated in \fref{fig:radial_strings}. All of this suggests that a braneworld observer can interpret the volume divergence appearing in the context of renormalization of $G_4$ in section \ref{sec:in-out}, as a mass renormalization of the effective particles induced on her world. 
    \begin{figure}[t]
   	\begin{subfigure}[b]{0.47\textwidth}
   		\def\svgwidth{\linewidth}
   		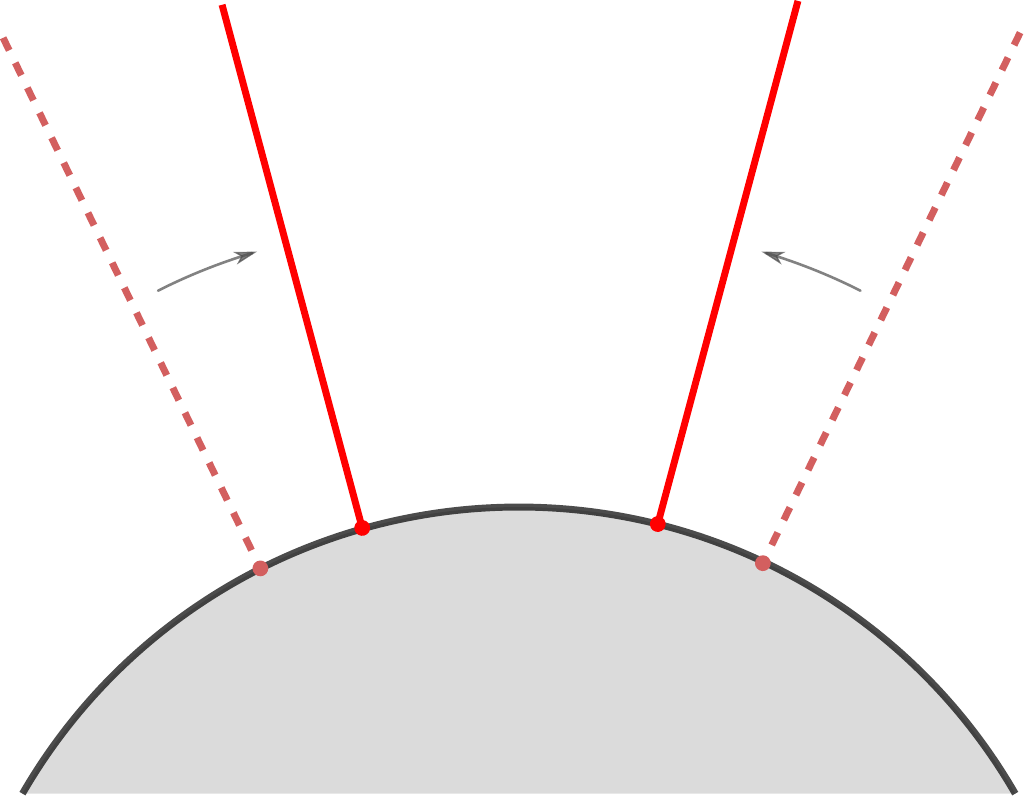
   		\caption{}
   		\label{fig:radial_strings1}
   	\end{subfigure}
   	~
   	\begin{subfigure}[b]{0.47\textwidth}
   		\def\svgwidth{\linewidth}
   		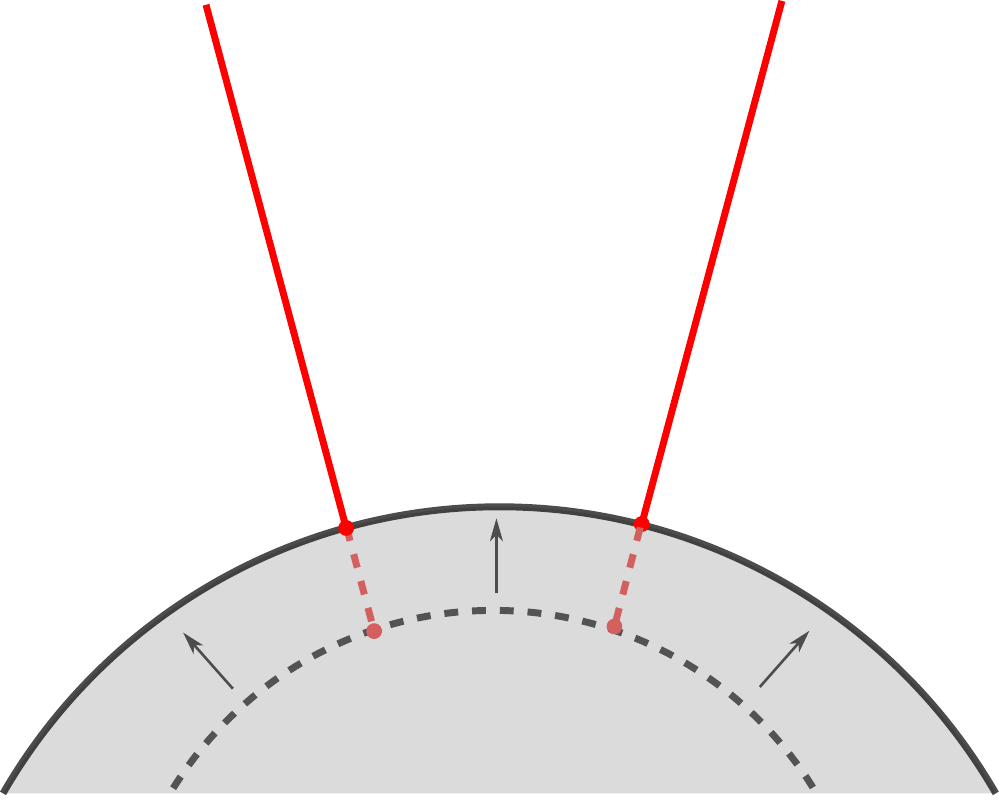
   		\caption{}
   		\label{fig:radial_strings2}
   	\end{subfigure}
   	\caption{Effect of gravity and expansion of the shell on four dimensional matter. Radially stretched strings gravitate towards each other while remaining radial, as shown in (a). Their end points trace four dimensional geodesics, behaving like matter particles on the shell. In the limit where the gravitational attraction between them can be ignored, the expanding shell ``eats up'' the strings and the four dimensional matter represented by their end points seem to move away from each other, corresponding to the expansion of the universe. This is shown in (b).}
   	\label{fig:radial_strings}
   \end{figure}
    
    \subsection{Mass renormalization = holographic renormalization}
    
    Interestingly, the mass renormalization can also be interpreted in the light of 
    holographic renormalization \cite{deHaro:2000vlm, Bianchi:2001de, Bianchi:2001kw, Skenderis:2002wp}. Physically, this  becomes very natural if we think of the extra dimension such that each radial position of the brane can be associated with an intrinsic energy scale. This scale can be identified as the calibration that an experimentalist living on a particular radial slice uses to define an effective low energy physics on her universe living on that slice. This physical idea is technically nurtured in the AdS/CFT correspondence \cite{Maldacena:1997re, Witten:1998qj, Gubser:1998bc}.
    
    Using standard techniques, the divergence due to infinite extra dimension can be easily tamed. Since the boundary of AdS is a special hypersurface on which a well defined field theory resides, all the physically measurable quantities on this particular hypersurface need to be finite. For us, this is applicable to the part of our spacetime that has access to the holographic boundary, namely the ``exterior AdS''. For this piece of AdS, one needs to add counterterms to the five dimensional bulk action of \fref{eq:R-EH}, which cancel potential IR divergences so that the new ``renormalized'' action renders all the physical quantities on the boundary finite.  The counterterm of interest is: 
    \be
    S_{\mbox{ct}}^{G_4} = -\kappa \int_{z=z_0} \D ^4 x \sqrt{\gamma^{+}} R^{\text{(4)}}[\gamma^{+}],
    \label{eq:Sct}
    \ee
    where $\gamma^{+}_{\mu\nu}$ is the induced metric on $z=z_0$ as a pullback of the exterior AdS. One should be able to choose a suitable $\kappa$ to get rid of the second term of \fref{eq:G45-G}. Note that the constant $\kappa$ is not an arbitrary choice for us. If we compute the boundary stress-energy tensor using the exterior AdS metric, we will find that this stress tensor will diverge. However, the same choice of $\kappa$ in the counterterm will also ensure a finite boundary energy.
    
    There is one catch. This argument on dimensional reduction and the subsequent renormalization using the particular counterterm of \fref{eq:Sct} relies on the fact that we have a non-flat metric on the cut-off boundary, $z = z_0$. In order to incorporate that, the five dimensional embedding needs to be spacetime dependent. This means, naively, that $z_0$ should also have a hidden dependence on the spacetime coordinates on the brane. In the context of a two brane model, when $z_0$ is finite, this would imply that $G_4$ would be a spacetime dependent quantity as argued in \cite{Montefalcone:2020vlu}.
    In particular, on a dS brane the effective $G_4$ would be time-dependent. Alternatively, one can trade off the spacetime dependence of $G_4$ at the price of introducing a five dimensional spacetime dependent bulk scalar field, the radion field, which can be interpreted as the distance between the two branes. This point of view was adopted in a recent work \cite{Karch:2020iit}.
    
    However, all of this can be avoided when the second brane is pushed all the way to the boundary of the AdS, $z_0 \rightarrow \infty$.\footnote{A more precise argument in connection to \cite{Karch:2020iit} will be presented in \fref{sec:time-dep}.} In this limit, the time-dependence of $z_0$ can be ignored, which, in the radion field interpretation, translates to the fact that the radion field scales with the cut-off as $\Phi_\textrm{rad} \sim 1/z_0$. With this in mind, we can use the renormalization scheme we have outlined to get an effectively constant $G_4$ given by
    \be
    \boxed{
        G_4 = 2{G_5 }\left(\frac{1}{\epsilon_- k_-} - \frac{1}{\epsilon_+ k_+}\right)^{-1}
    }
    \label{G4-renorm}
    \ee
    Let us quickly see an example of how such renormalization scheme actually works in typical situations relevant for the cases we will study here. Let us estimate \fref{eq:Sct} when $z_0 = z_0(t)$ and the induced metric takes the simple form 
    \be
    \D s^2 = e^{2 \, \epsilon_+ \, k_+ z_0(t)} \eta_{\mu\nu} \D x^\mu \D x^\nu .
    \ee
    For this metric, the Ricci scalar, $R^{\text{(4)}}[\gamma^{+}]$, scales as $e^{-2 \epsilon_+ k_+ z_0(t)}$, while $\sqrt{\gamma^{+}} $ scales as $e^{4 \epsilon_+ k_+ z_0(t)}$. This gives the correct exponential behavior required for the cancellation of the second term in \fref{eq:G45-G}. However, there will also be multiplicative factors proportional to first and second derivatives of $z_0(t)$, and one needs to integrate all of these over time. However, in the limit $z_0(t) \rightarrow \infty$ one can neglect the contribution from the spatial curvature and simply consider exponential expansion of the brane with  $z_0(t) = H_0 t/(\epsilon_+ k_+)$. In this case, the integration simply yields the desired cancellation upon a suitable choice of the constant, $\kappa$.
    
    One can also consider global AdS spacetimes separated by spherical branes, with induced metrics of FLRW form, where we actually know the spacetime dependence of the embedding function. This set up is close to what we actually considered in our model of dark bubble. Here one can compute the counterterm precisely, following existing holographic literature \cite{Banerjee:2012dw} \footnote{Here we presented a heuristic argument assuming one can keep the cutoff sufficiently close to the boundary so that the standard holographic renormalization still applies. This approach is not strictly correct since our brane is always sitting at a finite cutoff. For an exact holographic derivation of $G_4$ one will have to resort to the holographic dual of cutoff AdS, namely the $TT$ deformed CFT \cite{McGough:2016lol, Hartman:2018tkw, Taylor:2018xcy, Guica:2019nzm}.}.
    
    \subsection{Curing signusitis}
    \noindent We also need to comment on the sign of the finite piece of $G_4$. The surprising sign, from the point of view of a naive dimensional reduction, is an important feature of the way four dimensional gravity is realized through the straight strings. This was an important part of the construction in \cite{Banerjee:2018qey} and was later elaborated in \cite{Banerjee:2020wix}, through an explicit computation of the graviton propagator on the shell. We noticed that in order to have the correct graviton propagator on the shell we need to add a non-normalizable piece in the graviton solution in AdS. The strategy of finding the propagator on the shell was as follows.
    \begin{itemize}
        \item We focused on the transverse traceless mode, for which the scalar propagator essentially has all the relevant information about the graviton propagator. For simplicity we worked in the Poincare patch of the AdS spaces.
        
        \item We first solved the equation for the scalar Green's function away from the source so that the solutions are given as linear combinations of the Bessel functions $K_2(p z/k_\pm)$ and $I_2(p z/k_\pm)$, $p$ being the transverse momentum. In the interior, we needed to discard $I_2(p z/k_-)$ to ensure regularity at the Poincare horizon. All together we had $3$ constants to be fixed, which were associated with the $K_2$ and $I_2$ parts of the solution in the exterior and with $I_2$ in the interior.
        
        \item We fixed two of the three constants by invoking the thin-shell junction conditions across the shell, which in the scalar case translates into the continuity of the fields and discontinuity of their derivatives across the brane source.
        
        \item Fixing the remaining constant carries all the crucial difference. When we face a similar question in holography with no source at the boundary, as would be the case for two pure AdS$_5$ spacetimes on two sides of the brane, we would simply impose a ``normalizability'' condition. In other words, the field would smoothly decay as a function of the radial coordinate as it goes to the boundary of AdS. However, for our construction to reproduce the correct effective graviton propagator on the brane, we need to do exactly the opposite, picking up the non-normalizable contribution while throwing away its normalizable counterpart. This gives the correct effective four dimensional Newton's constant on the brane with the desired sign and also yields the correct graviton propagator at low energy. 
    \end{itemize}
    The natural interpretation of this non-normalizable mode is simply that we cannot afford to have pure AdS spacetimes on either side of the brane. Instead, there must exist an extended stringy ``non-normalizable'' source in the bulk that ensures the correct sign of the effective Newton's constant as well as the correct behavior of the graviton propagator on the shell. As explained in \cite{Banerjee:2020wix}, it is easy to understand how this works by looking at how a string pulls on the brane in \fref{fig:brane_bending}.
    \begin{figure}[t]
        \centering
        \def\svgwidth{\linewidth}
\begingroup%
  \makeatletter%
  \providecommand\color[2][]{%
    \errmessage{(Inkscape) Color is used for the text in Inkscape, but the package 'color.sty' is not loaded}%
    \renewcommand\color[2][]{}%
  }%
  \providecommand\transparent[1]{%
    \errmessage{(Inkscape) Transparency is used (non-zero) for the text in Inkscape, but the package 'transparent.sty' is not loaded}%
    \renewcommand\transparent[1]{}%
  }%
  \providecommand\rotatebox[2]{#2}%
  \newcommand*\fsize{\dimexpr\f@size pt\relax}%
  \newcommand*\lineheight[1]{\fontsize{\fsize}{#1\fsize}\selectfont}%
  \ifx\svgwidth\undefined%
    \setlength{\unitlength}{1150.17174847bp}%
    \ifx\svgscale\undefined%
      \relax%
    \else%
      \setlength{\unitlength}{\unitlength * \real{\svgscale}}%
    \fi%
  \else%
    \setlength{\unitlength}{\svgwidth}%
  \fi%
  \global\let\svgwidth\undefined%
  \global\let\svgscale\undefined%
  \makeatother%
  \begin{picture}(1,0.41340215)%
    \lineheight{1}%
    \setlength\tabcolsep{0pt}%
    \put(0,0){\includegraphics[width=\unitlength,page=1]{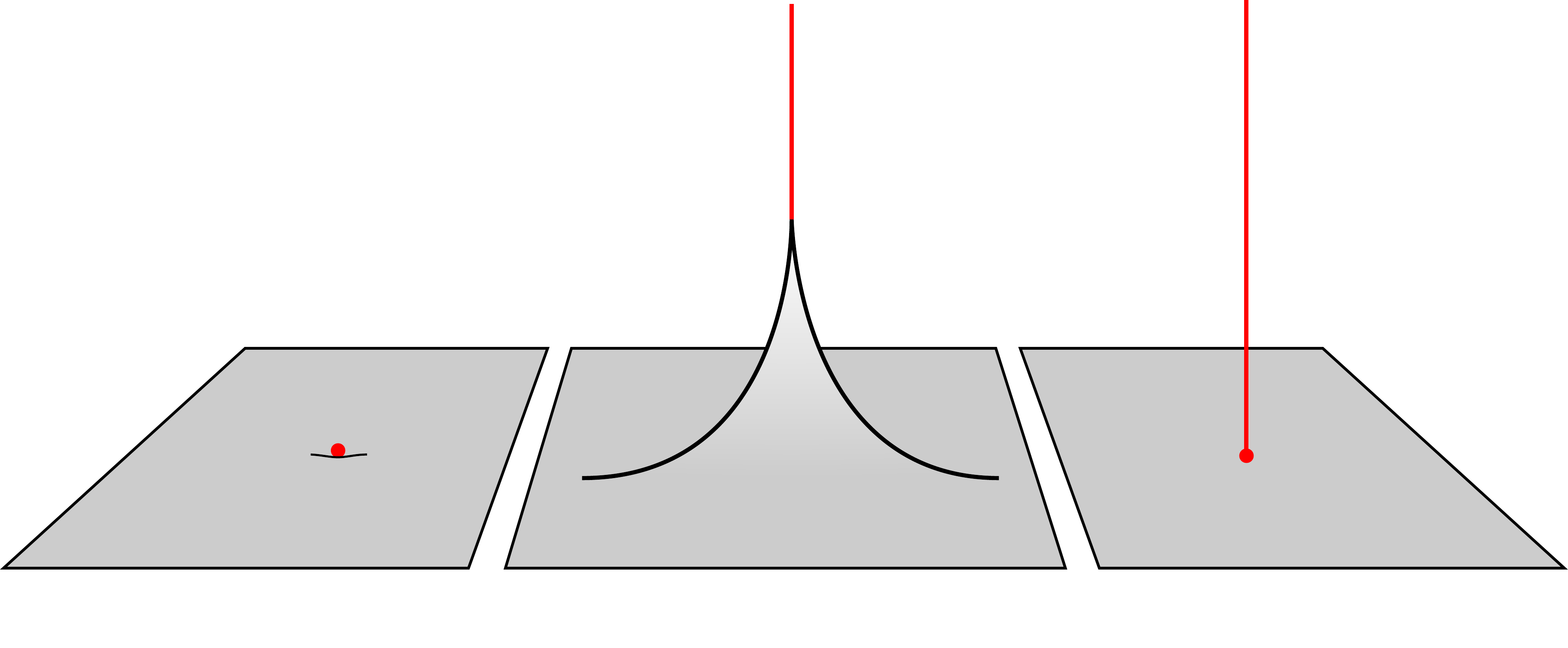}}%
    \put(0.30438751,0.24928642){\color[rgb]{0,0,0}\makebox(0,0)[lt]{\lineheight{1.25}\smash{\begin{tabular}[t]{l}$k_+$\end{tabular}}}}%
    \put(-0.12524937,0.20261918){\color[rgb]{0,0,0}\makebox(0,0)[lt]{\begin{minipage}{0.06455227\unitlength}\raggedright \end{minipage}}}%
    \put(0.48915178,0.00612739){\color[rgb]{0,0,0}\makebox(0,0)[lt]{\lineheight{1.25}\smash{\begin{tabular}[t]{l}$k_-$\end{tabular}}}}%
    \put(0,0){\includegraphics[width=\unitlength,page=2]{brane_bending2.pdf}}%
    \put(0.05050419,0.28698063){\color[rgb]{0,0,0}\makebox(0,0)[lt]{\lineheight{1.25}\smash{\begin{tabular}[t]{l}$\xi$\end{tabular}}}}%
  \end{picture}%
\endgroup%

        \caption{Bending of the brane in response to a matter source placed on it. A matter source placed directly on the brane (as shown in the first panel) bends it ``downwards'', giving a negative mass object in four dimensions. A string pulling upwards on the brane (second panel) causes the brane to bend ``upwards'' \ie, in the direction of increasing $\xi$ and results in a positive mass in four dimensions. Using a gauge transformation, this can be brought into an ``unbent'' gauge where the brane appears flat and the four dimensional matter source is the end point of the stretched string (as shown in the third panel).}
 \label{fig:brane_bending}
    \end{figure}
    One can show that a string with constant tension $\tau$ stretching in the radial direction, $z$ originating from the brane located at $\xi = 0$ and extending up to the radial position $\xi_0$, yields two different types of effects -- a purely bulk effect due to the presence of the string and a localized effect on the brane, which can be interpreted as a massive deformation in the four dimensional theory. This can be easily seen by computing the energy density of such a configuration and imposing the covariant conservation of the energy momentum tensor.
    \begin{equation}\label{eq:tauoverk}
    \epsilon=\tau\frac{1}{a(\xi)^3}\delta\left(x^a-x^a_{0}\right)\Theta\left(\xi-\xi_0\right)
    +\frac{\tau}{k}\frac{1}{a(\xi)^3}\delta\left(x^a-x^a_{0}\right)\delta\left(\xi-\xi_0\right).
    \end{equation}
    While the first term is the bulk effect due to the extended string, the second term denotes effective energy density of a point particle with mass $\tau/ k$ localized on the brane. Here $a(\xi)$ is the scale factor appearing in the induced metric of the brane as
    \be
    \D s^2 = \D z^2 + a(\xi)^2\eta_{ab}\D x^a \D x^b,
    \ee
    and $x^a_{0}$ is the transverse position of the string.
    
    As we see from the figure, a string stretching upwards will pull on the brane. As explained in \cite{Banerjee:2020wix}, the induced metric on the bent brane in the perturbed background will give four dimensional gravity. If we choose a gauge in which the brane is straight, we need to add a mass to the endpoint of the string that pulls the string down. As we showed in \fref{eq:tauoverk}, the required mass is $\tau/ k$, which is also what a four dimensional observer measures. Looking at the figure, we see that the effect of the string is exactly opposite of that of matter lying directly on the brane. Such matter will pull the brane down, and effectively work as a negative energy. This explains the sign of $G_4$.
    
    \section{Crouching strings - hidden time}
       \label{sec:time-dep}
   In this section, we will comment on some recent work \cite{Montefalcone:2020vlu, Karch:2020iit} and its connection to our model of a dark bubble. In particular, we will address the issue of residual time-dependence.
   In \cite{Karch:2020iit}, the authors considered a generalized two brane model with ``mismatching'' tension in order to have a time evolving de Sitter braneworld. They showed that one can obtain de Sitter in the Einstein frame with a time independent $G_4$. As they point out, however, there still remains a time-dependence due to explicit time-dependence in the radion field. If the second brane is pushed to infinity, the contribution from the radion will completely decouple from the gravity sector,  but matter residíng on the IR-brane will still inherit the time dependence through coupling constants depending on the radion. The authors in \cite{Montefalcone:2020vlu} arrived at a very similar conclusion, although from a more general argument based on dimensional reduction. In our case this time dependence will not appear since matter is introduced in a very different way.
   
   There are two major differences in our model. The first difference is that our bubble has an inside and an outside. As for the time-dependence, this consideration by itself will not make much of a difference. If we study the effective four dimensional cosmological constant as a function of the brane tension, as summed up in \fref{fig:4dCC-sigma}, one concludes that the single brane limit of \cite{Karch:2020iit} lies close to the critical tension given by $k_- +k_+$, while our model is close to $k_- -k_+$. 
   
   The second and the more important difference is that in order to support the inside-outside configuration to give a well defined gravity on the brane, one would need to have non-local extended sources in the bulk in the form of radially stretched strings. We have also seen that a string in the bulk imprints matter on the brane on which it ends. The end points of the strings attached to the brane are secretly time-dependent since the brane is moving with its radial position being a function of time. In the language of \cite{Karch:2020iit}, these end points would correspond to localized matter fields on the brane and hence will be time-dependent in exactly the same dynamical way prescribed in their paper. 
   \begin{figure}[t]
       \centering
       \def\svgwidth{0.9\linewidth}
       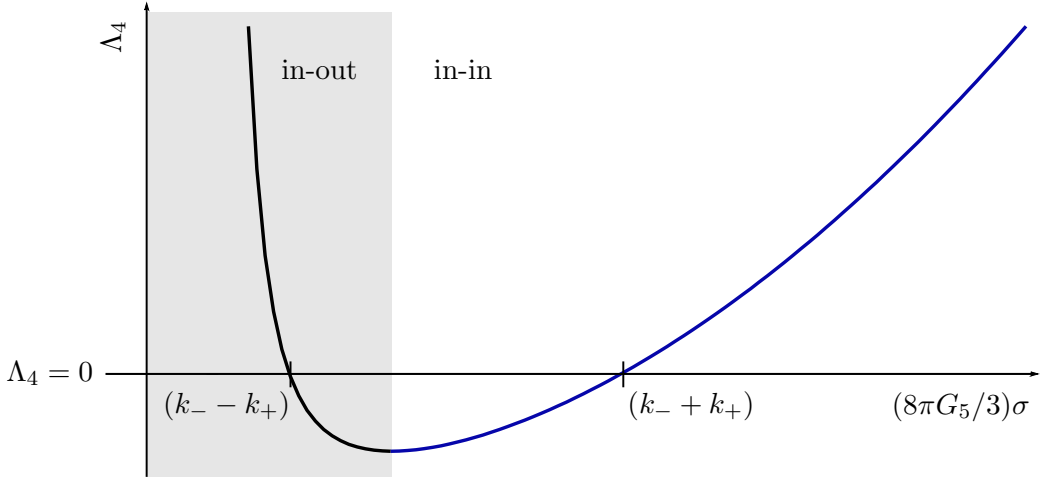
       \caption{The four dimensional cosmological constant on the brane as a function of its tension. The curve in the shaded region on the left represents the shellworld (inside-outside), while the one to the right is the RS braneworld (inside-inside). The tension corresponding to a Minkowski brane are represented on the horizontal axis.}
       \label{fig:4dCC-sigma}
   \end{figure}	
   However,  if we take into account the dynamics of the bulk strings as considered in \fref{sec:mass-renorm}, the time dependence goes away. As we discussed there in detail, the dynamics of the bulk strings make sure that the mass is properly renormalized so that the measured mass on the four dimensional world is insensitive to the length of the five dimensional strings. This means the secret time dependence of radial embedding is also naturally taken care of, at least in the limit when there is no other brane close by. 
   
   So far we only talked about the single brane limit where the dynamics of the strings automatically takes care of the time dependence. One needs to be more careful in case there is a second brane at a finite distance from our braneworld.  However, we would like to emphasize that the two brane model of RS \cite{Randall:1999ee} was originally proposed to incorporate the standard model of particles in the braneworld scenario which, when generalized for the mismatched configuration in \cite{Karch:2020iit}, suffers from the undesired time dependence. Our model, on the other hand, does not need a second brane -- matter fields are automatically introduced by the end points of strings. As we already saw in \cite{Banerjee:2019fzz}, our model has the potential to describe both gravity and particle physics on the same brane. A more detailed study to realize the standard model on particle physics on the shellworld is in progress.

    \section{Summary}
    \label{sec:discussion}
    To summarize, in this article we have emphasized how the radially stretched strings in the shellworld construction give rise to a finite four dimensional Newton's constant, despite having a five dimensional AdS spacetime with infinite volume. By performing a world sheet analysis of a radially stretched probe string moving in the background of a heavy string, we have shown that the end point of the probe string follows a four dimensional geodesic, mimicking a gravitating four dimensional massive particle on the shellworld. This is consistent with the effective four dimensional gravity obtained in the shellworld construction. Together with the \emph{brane bending} analysis performed in \cite{Banerjee:2020wix}, this serves to further highlight the fact that four dimensional matter on the shellworld needs to be realized not as localized matter fields on the brane, but rather as radially outstretched strings. Finally, we have commented on two recent results which point out a time dependence that arises in braneworld models where matter is localized on the brane. However, the outstretched strings which give rise to four dimensional matter on the shellworld, precisely make sure that there is no such time dependence in these constructions. 
	
	It is true that matter on the brane naively contributes with a negative sign to the effective energy density. But from the four dimensional perspective one always needs to take into account gravitational contributions as well, in order to obtain the effective four dimensional energy density. As long as our model can be considered to be part of a higher dimensional consistent theory of quantum gravity, such as string theory, there cannot be any issues. The matter on the brane is described by a brane action with a positive kinetic term. This action is coupled to gravity and induces, in particular through the junction conditions, the physics we have discussed in the limits stated.
		
	If one wants to construct a four dimensional field theory it is clear that one must combine the brane degrees of freedom with higher dimensional gravity. Doing this carefully, one should also be able to see how this field theory breaks down in certain limits. This is an interesting problem that remains. One could certainly expect new kinds of phenomena to appear in this context, but, as we have argued, no inconsistencies should appear.

    \section*{Acknowledgments}
    The work of S.B. is supported by the Alexander von Humboldt postdoctoral fellowship. We would like to thank the referee for invaluable comments and discussion.

    \small
    \bibliography{mainv4}

\providecommand{\href}[2]{#2}\begingroup\raggedright\begin{thebibliography}{10}

\bibitem{Banerjee:2018qey}
S.~Banerjee, U.~Danielsson, G.~Dibitetto, S.~Giri and M.~Schillo,  {\em
  {Emergent de Sitter Cosmology from Decaying Anti–de Sitter Space}}, Phys.
  Rev. Lett. {\bf 121} (2018), no.~26, 261301
[\href{http://www.arXiv.org/abs/1807.01570}{{\tt 1807.01570}}].

\bibitem{Banerjee:2019fzz}
S.~Banerjee, U.~Danielsson, G.~Dibitetto, S.~Giri and M.~Schillo,  {\em {de
  Sitter Cosmology on an expanding bubble}}, JHEP {\bf 10} (2019) 164
[\href{http://www.arXiv.org/abs/1907.04268}{{\tt 1907.04268}}].

\bibitem{Banerjee:2020wix}
S.~Banerjee, U.~Danielsson and S.~Giri,  {\em {Dark bubbles: decorating the
  wall}}, JHEP {\bf 04} (2020) 085
  [\href{http://www.arXiv.org/abs/2001.07433}{{\tt 2001.07433}}].

\bibitem{Danielsson:2018ztv}
U.~H. Danielsson and T.~Van~Riet,  {\em {What if string theory has no de Sitter
  vacua?}}, Int. J. Mod. Phys. {\bf D27} (2018), no.~12, 1830007
[\href{http://www.arXiv.org/abs/1804.01120}{{\tt 1804.01120}}].

\bibitem{Cicoli:2018kdo}
M.~Cicoli, S.~De~Alwis, A.~Maharana, F.~Muia and F.~Quevedo,  {\em {De Sitter
  vs Quintessence in String Theory}}, Fortsch. Phys. {\bf 67} (2019), no.~1-2,
  1800079 [\href{http://www.arXiv.org/abs/1808.08967}{{\tt 1808.08967}}].

\bibitem{Brennan:2017rbf}
T.~D. Brennan, F.~Carta and C.~Vafa,  {\em {The String Landscape, the
  Swampland, and the Missing Corner}}, PoS {\bf TASI2017} (2017) 015
  [\href{http://www.arXiv.org/abs/1711.00864}{{\tt 1711.00864}}].

\bibitem{Palti:2019pca}
E.~Palti,  {\em {The Swampland: Introduction and Review}}, Fortsch. Phys. {\bf
  67} (2019), no.~6, 1900037 [\href{http://www.arXiv.org/abs/1903.06239}{{\tt
  1903.06239}}].

\bibitem{Koga:2019yzj}
I.~Koga and Y.~Ookouchi,  {\em {Catalytic Creation of Baby Bubble Universe with
  Small Positive Cosmological Constant}}, JHEP {\bf 10} (2019) 281
[\href{http://www.arXiv.org/abs/1909.03014}{{\tt 1909.03014}}].

\bibitem{Basile:2020mpt}
I.~Basile and S.~Lanza,  {\em {de Sitter in non-supersymmetric string theories:
  no-go theorems and brane-worlds}},
  \href{http://www.arXiv.org/abs/2007.13757}{{\tt 2007.13757}}.

\bibitem{Randall:1999vf}
L.~Randall and R.~Sundrum,  {\em {An Alternative to compactification}}, Phys.
  Rev. Lett. {\bf 83} (1999) 4690--4693
[\href{http://www.arXiv.org/abs/hep-th/9906064}{{\tt hep-th/9906064}}].

\bibitem{Montefalcone:2020vlu}
G.~Montefalcone, P.~J. Steinhardt and D.~H. Wesley,  {\em {Dark energy, extra
  dimensions, and the Swampland}}, JHEP {\bf 06} (2020) 091
  [\href{http://www.arXiv.org/abs/2005.01143}{{\tt 2005.01143}}].

\bibitem{Karch:2020iit}
A.~Karch and L.~Randall,  {\em {Geometries with mismatched branes}},
  \href{http://www.arXiv.org/abs/2006.10061}{{\tt 2006.10061}}.

\bibitem{Dibitetto:2020csn}
G.~Dibitetto, N.~Petri and M.~Schillo,  {\em {Nothing really matters}}, JHEP
  {\bf 08} (2020) 040 [\href{http://www.arXiv.org/abs/2002.01764}{{\tt
  2002.01764}}].

\bibitem{GarciaEtxebarria:2020xsr}
I.~García~Etxebarria, M.~Montero, K.~Sousa and I.~Valenzuela,  {\em {Nothing
  is certain in string compactifications}},
  \href{http://www.arXiv.org/abs/2005.06494}{{\tt 2005.06494}}.

\bibitem{Israel:1966rt}
W.~Israel,  {\em {Singular hypersurfaces and thin shells in general
  relativity}}, Nuovo Cim. {\bf B44S10} (1966) 1
[Erratum: Nuovo Cim.B48,463(1967); Nuovo Cim.B44,1(1966)].

\bibitem{doi:10.1002/andp.19243791403}
K.~Lanczos,  {\em Flächenhafte Verteilung der Materie in der Einsteinschen
  Gravitationstheorie}, Annalen der Physik {\bf 379} (1924), no.~14, 518--540
  [\href{http://www.arXiv.org/abs/https://onlinelibrary.wiley.com/doi/pdf/10.1002/andp.19243791403}{{\tt
  https://onlinelibrary.wiley.com/doi/pdf/10.1002/andp.19243791403}}].

\bibitem{doi:10.1002/andp.19243780505}
N.~Sen,  {\em Über die Grenzbedingungen des Schwerefeldes an
  Unstetigkeitsflächen}, Annalen der Physik {\bf 378} (1924), no.~5‐6,
  365--396
  [\href{http://www.arXiv.org/abs/https://onlinelibrary.wiley.com/doi/pdf/10.1002/andp.19243780505}{{\tt
  https://onlinelibrary.wiley.com/doi/pdf/10.1002/andp.19243780505}}].

\bibitem{Garriga:1999yh}
J.~Garriga and T.~Tanaka,  {\em {Gravity in the brane world}}, Phys. Rev. Lett.
  {\bf 84} (2000) 2778--2781
[\href{http://www.arXiv.org/abs/hep-th/9911055}{{\tt hep-th/9911055}}].

\bibitem{Giddings:2000mu}
S.~B. Giddings, E.~Katz and L.~Randall,  {\em {Linearized gravity in brane
  backgrounds}}, JHEP {\bf 03} (2000) 023
[\href{http://www.arXiv.org/abs/hep-th/0002091}{{\tt hep-th/0002091}}].

\bibitem{deHaro:2000vlm}
S.~de~Haro, S.~N. Solodukhin and K.~Skenderis,  {\em {Holographic
  reconstruction of space-time and renormalization in the AdS / CFT
  correspondence}}, Commun. Math. Phys. {\bf 217} (2001) 595--622
[\href{http://www.arXiv.org/abs/hep-th/0002230}{{\tt hep-th/0002230}}].

\bibitem{Bianchi:2001de}
M.~Bianchi, D.~Z. Freedman and K.~Skenderis,  {\em {How to go with an RG
  flow}}, JHEP {\bf 08} (2001) 041
  [\href{http://www.arXiv.org/abs/hep-th/0105276}{{\tt hep-th/0105276}}].

\bibitem{Bianchi:2001kw}
M.~Bianchi, D.~Z. Freedman and K.~Skenderis,  {\em {Holographic
  renormalization}}, Nucl. Phys. B {\bf 631} (2002) 159--194
  [\href{http://www.arXiv.org/abs/hep-th/0112119}{{\tt hep-th/0112119}}].

\bibitem{Skenderis:2002wp}
K.~Skenderis,  {\em {Lecture notes on holographic renormalization}}, Class.
  Quant. Grav. {\bf 19} (2002) 5849--5876
  [\href{http://www.arXiv.org/abs/hep-th/0209067}{{\tt hep-th/0209067}}].

\bibitem{Maldacena:1997re}
J.~M. Maldacena,  {\em {The Large N limit of superconformal field theories and
  supergravity}}, Int. J. Theor. Phys. {\bf 38} (1999) 1113--1133
  [\href{http://www.arXiv.org/abs/hep-th/9711200}{{\tt hep-th/9711200}}],
[Adv. Theor. Math. Phys.2,231(1998)].

\bibitem{Witten:1998qj}
E.~Witten,  {\em {Anti-de Sitter space and holography}}, Adv. Theor. Math.
  Phys. {\bf 2} (1998) 253--291
[\href{http://www.arXiv.org/abs/hep-th/9802150}{{\tt hep-th/9802150}}].

\bibitem{Gubser:1998bc}
S.~S. Gubser, I.~R. Klebanov and A.~M. Polyakov,  {\em {Gauge theory
  correlators from noncritical string theory}}, Phys. Lett. {\bf B428} (1998)
  105--114
[\href{http://www.arXiv.org/abs/hep-th/9802109}{{\tt hep-th/9802109}}].

\bibitem{Banerjee:2012dw}
S.~Banerjee, S.~Bhowmick, A.~Sahay and G.~Siopsis,  {\em {Generalized
  Holographic Cosmology}}, Class. Quant. Grav. {\bf 30} (2013) 075022
  [\href{http://www.arXiv.org/abs/1207.2983}{{\tt 1207.2983}}].

\bibitem{McGough:2016lol}
L.~McGough, M.~Mezei and H.~Verlinde,  {\em {Moving the CFT into the bulk with
  $ T\overline{T} $}}, JHEP {\bf 04} (2018) 010
[\href{http://www.arXiv.org/abs/1611.03470}{{\tt 1611.03470}}].

\bibitem{Hartman:2018tkw}
T.~Hartman, J.~Kruthoff, E.~Shaghoulian and A.~Tajdini,  {\em {Holography at
  finite cutoff with a $T^2$ deformation}}, JHEP {\bf 03} (2019) 004
[\href{http://www.arXiv.org/abs/1807.11401}{{\tt 1807.11401}}].

\bibitem{Taylor:2018xcy}
M.~Taylor,  {\em {TT deformations in general dimensions}},
\href{http://www.arXiv.org/abs/1805.10287}{{\tt 1805.10287}}.

\bibitem{Guica:2019nzm}
M.~Guica and R.~Monten,  {\em {$T\bar T$ and the mirage of a bulk cutoff}},
  \href{http://www.arXiv.org/abs/1906.11251}{{\tt 1906.11251}}.

\bibitem{Randall:1999ee}
L.~Randall and R.~Sundrum,  {\em {A Large mass hierarchy from a small extra
  dimension}}, Phys. Rev. Lett. {\bf 83} (1999) 3370--3373
[\href{http://www.arXiv.org/abs/hep-ph/9905221}{{\tt hep-ph/9905221}}].

\end{thebibliography}\endgroup
    \bibliographystyle{utphysmodb}
\end{document}